\documentclass[useAMS,usenatbib,usegraphics]{mn2e}
\usepackage{graphicx,epsfig}  
\usepackage{amsmath,amssymb}
\def\ut#1{\mathop{\vtop{\ialign{##\crcr
     $\hfil\displaystyle{#1}\hfil$\crcr\noalign
     {\kern1pt\nointerlineskip}\hbox{$\hfil\sim\hfil$}\crcr
     \noalign{\kern1pt}}}}}

\def\undersymbol#1#2{\mathop{\vtop{\ialign{##\crcr
     $\hfil\displaystyle{#2}\hfil$\crcr\noalign
     {\kern1pt\nointerlineskip}\hbox{$\hfil#1\hfil$}\crcr
     \noalign{\kern1pt}}}}}

\title[Polarization in microlensing events towards the Galactic bulge]
{Polarization in microlensing towards the Galactic bulge}
\author[G. Ingrosso et al.]
{G. Ingrosso,$^{1,2}$\thanks{E-mail:
ingrosso@le.infn.it}
 S. Calchi Novati,$^{3,4}$
 F. De Paolis,$^{1,2}$
\newauthor Ph. Jetzer,$^5$
A. A. Nucita,$^{1,2}$ 
F. Strafella$^1$  and
A. F. Zakharov$^{6,7}$\\
$^1$  Dipartimento di Matematica e Fisica, ``Ennio De Giorgi'', 
Universit\`a del Salento, CP 193, I-73100 Lecce, Italy \\
$^2$   {\it INFN} Sezione di Lecce, CP 193, I-73100 Lecce, Italy \\
$^3$ Dipartimento di Fisica ``E. R. Caianiello'', Universit\`a di 
Salerno,
Via Ponte don Melillo, 84084 Fisciano (SA), Italy\\
$4$ Istituto Internazionale per gli Alti Studi Scientifici (IIASS), 
Vietri Sul Mare (SA), Italy\\
$^5$       Institute for Theoretical Physics,
           University of  Z\"{u}rich, Winterthurerstrasse 190,
           CH-8057 Z\"{u}rich, Switzerland\\
$^6$       Institute of Theoretical and Experimental Physics,
            B. Cheremushkinskaya 25, 117259 Moscow,  Russia \\
$^7$      Bogoliubov Laboratory of Theoretical Physics, Joint Institute for Nuclear Research, 141980 Dubna, Russia}
\begin{document}

\date{Accepted xxx; Received xxx;
in original form xxx}

\pagerange{\pageref{firstpage}--\pageref{lastpage}} \pubyear{2011}

\maketitle

\label{firstpage}

\begin{abstract}

Gravitational microlensing, when finite size source effects are relevant,
provides an unique tool for the study of source star stellar atmospheres
through an enhancement of a 
characteristic polarization signal.
This is due to the differential magnification induced 
during the crossing of the source star.
In this paper we consider a specific set of reported
highly magnified, both single and binary exoplanetary systems,
microlensing events towards the Galactic bulge and evaluate the expected 
polarization signal. To this purpose,
we consider several polarization models which apply
to different types of source stars: { hot, late type main sequence
and cool giants}.  As a result we compute the  polarization signal $P$,
which goes up
to P=0.04 { percent} for { late type} stars and up to a few percent
for cool giants, depending on the underlying
physical polarization processes and atmosphere model parameters.
Given a $I$ band magnitude at maximum
magnification of about  12, and a typical duration
of the polarization signal up to 1 day, we
conclude that the currently available  technology,
in particular the polarimeter in FORS2 on the VLT, 
potentially may allow the detection of such signals.
This observational programme may take advantage
of the currently available surveys plus follow up
strategy already routinely used for microlensing monitoring
towards the Galactic bulge (aimed at the detection
of exoplanets). In particular, this allows one to predict in advance
for which events and at which exact time the
observing resources may be focused to make intensive
polarization measurements.

\end{abstract}


\section{Introduction}

Gravitational microlensing, initially developed to search 
for MACHOs in { the} Galactic halo and near the Galactic disc 
\citep{Pacz86,Macho93,Eros93,Ogle93,Ogle94} 
has become nowadays a powerful tool to 
investigate several aspects of stellar 
astrophysics and also to search for extrasolar planets orbiting around lens
stars.

Indeed, microlensing gives the opportunity to study the star's limb-darkening
profile, which is the variation of the intensity 
from the disc center to the limb,
and thus to test stellar atmosphere models. 
At the same time, microlensing leads to the discovery and the detailed  
characterization of exoplanetary systems when planetary deviations in the 
light-curves expected for single-lens events are detected
(see \cite{Dominik10} and \cite{Gaudi10} for recent reviews). 

Microlenses can spatially resolve a source star thanks to caustic structures 
created by a lens system \citep{SEF}. 
{ Caustics are formed by a set of closed curves,}  along which the 
point source magnification is formally infinite, with a steep increase in 
magnification in their vicinity. This increase is so steep that the 
characteristic length scale of the differential magnification effect is of 
the order of a fraction of the source star radius. In this way different 
parts of the source star are magnified by substantially different amounts.
The resulting lensing light-curve deviates from the standard form expected for
a point source event \citep{Witt94,Gould94,Alcock97}  and the analysis of the 
deviations enables to measure the limb-darkening profile of the lensed star.

Early works \citep{Witt95,Loeb-Sasselov95,Valls-Gabaud98,Heyrovsky03} 
have pointed out the sensitivity of microlensing light-curves
to limb-darkening, with the aim to help to remove the microlensing model 
parameter degeneracy. 
The specific use of microlensing as a tool to study stellar 
atmospheres was proposed later \citep{Hendry98,Gaudi-Gould99}, 
in particular to probe atmospheres of red giants in the Galactic bulge  
\citep{Heyrovsky00}. 

{The best candidate events for studying stellar atmospheres are 
     highly magnified microlensing events, which also show
     relevant finite size source effects. For these events  
     the lens and the background source star are almost 
     aligned and the lens passes over the surface of the source star 
     ({\it transit} events). 
Although relatively rare, these events  
potentially contain unique information 
on the stellar atmosphere properties of the source star as shown by 
\cite{Fouque10} and \cite{Zub10}.
Indeed, besides the brightness profile of a remote source star disc, 
highly magnified events with large finite size source effects   
allow to measure the lens Einstein radius $R_{\rm E}$ 
(if the physical radius $R_*$ of the source is known) 
and provide a unique chance to study spectroscopically Galactic bulge stars.

The light-curve analysis of highly magnified events is also sensitive 
to the presence of lens planetary companions, in particular when
the planet-to star distance is of the order of $R_{\rm E}$.
The same opportunity for studying stellar atmospheres is offered by 
binary microlensing due to caustic crossing as
the source passes through fold \citep{SEF} and cusp caustics 
\citep{Schneider92,Zakharov95}.
}

The aim of the present paper is to consider polarization variability 
of the source star light for real events, taking into account 
different polarization  mechanisms 
according to the source star type.
Indeed, variations in the polarization curves are similar to finite source 
effects in microlensing when { color effects } may appear due to 
limb darkening and color distribution across the disc 
\citep{Witt94,Bogdanov95a,Bogdanov95b,
Gaudi-Gould99,Bogdanov00,Bogdanov02,Dominik05,Heyrovsky07}.

It is known that the light received from the stellar limb may
be significantly (up to about 12 { percent}) polarized due to the so called
Sobolev -- Chandrasekhar effect \citep{Sobolev49,Chandra60,Sobolev63}. 
Polarization parallel to the limb of the source is caused by photon 
(Thomson) scattering off free electrons, when the light passes through the 
stellar atmosphere. 
This polarization mechanism is effective 
{ for hot stars of any luminosity class}
which, indeed, have a free electron atmosphere.

It is also known that, by a minor extent, the continuum spectrum of the 
{  main sequence stars of late type} is linearly polarized by coherent 
(Rayleigh) scattering on neutral hydrogen in its ground state and, with 
a minor contribution, by scattering on free electrons \citep{Stenflo99}. 
The polarization in { late type stars} has been measured only for the 
Sun (for which
due to the distance the projected disc is spatially resolved) and 
also in this case { as for hot stars}, the polarization 
gets its maximum value near the 
solar limb, due to the most favorable geometry there. 

However, the light received from the stars is usually unpolarized, 
{ since the flux from each stellar disc element is the same.}
A net polarization of the stellar light may be introduced by some
suitable asymmetry in the stellar disc (e.g. eclipses, tidal distortions, 
stellar spots, fast rotation, magnetic fields) and also in the propagation 
through the interstellar medium \citep{BKS85,Dolginov95}.

In the microlensing context, polarization in the stellar light is induced  
due to the proper motion of the lens star through the source star
disc, during which different parts are magnified
by different amounts.
Therefore, the gravitational lens scans the disc of the background star
giving rise not only to a time dependent gravitational magnification 
of the source star light but also to a time dependent polarization.

As for { the} limb-darkening brightness profile, 
the polarization degree is maximized when the source trajectory crosses
regions with high magnification gradient. This occurs in
{\it transit} events since the magnification of the source star flux 
increases as the lens approaches closer to the source star, 
and during binary microlensing events since now there is the 
opportunity for caustic crossing.

Accordingly, we consider the polarization variability for the 
{\it transit} events (with lenses passing over source stars) selected 
by \cite{Choi} and
for a subset of exoplanetary events towards the Galactic 
bulge \citep{Gaudi10}.
{ Exoplanetary events are binary lens systems characterized by values
of the planet-to-star mass ratio $q \ll 1$ and have smaller star-to-planet 
distance
$d$ as compared to the separation of the stars in a binary system.
A full treatment of the polarization in binary microlensing events
will be the subject of a subsequent paper.
Here, selecting a representative sample of exoplanetary events,
we show how the polarization works in binary microlensing.}

The idea that polarization induced by an electron scattering atmosphere 
could be enhanced by gravitational microlensing and eventually observed
was first raised and investigated in relation to supernovae by \cite{SW87}. 
In particular, the observation of a variable polarization  due to a
supernova beam expanding in a self-gravitating system constituted by 
individual masses ($ \ut > 10^{-3}~M_{\odot}$) may indicate the 
presence of dark matter objects.
    
Then, \cite{Simmons95a}, \cite{Simmons95b} and \cite{BCS96} 
pointed out that the polarization 
of star light induced by an electron scattering atmosphere 
is enhanced by the microlensing effect 
and that the relative motion of source and lens causes a time variation 
of the source polarization degree. 
\cite{Simmons95a} and \cite{Simmons95b} also present a numerical calculation 
of the polarization degree induced by a single-lens (the Schwarzschild lens), 
showing that during a microlensing event the polarization profile has a double
peak for {\it transit} events, (where a part of the source disc is aligned 
with the lens and observer), while it has a single peak in 
the {\it bypass} events (where the source trajectory remains outside the lens).

Assuming that the source star has an electron scattering atmosphere,
\cite{Agol96} calculated the time-dependent polarization of a star 
being gravitationally lensed by a binary system. 
Polarization as high as $P \simeq 1$ {percent} 
can be achieved if the star crosses a caustic or passes near a cusp; 
otherwise, the maximum polarization is $\simeq 0.1$ { percent}. 

Polarization by non-compact microlenses 
\citep{Gurevich95,Zakharov96a,Zakharov96b,Zakharov99,Zakharov10}
was also analyzed \citep{Belokurov98,Zakharov98} since a source may cross
caustics arising in the model. 

The most likely candidates for observing polarization variability 
during microlensing events would be young, hot giant star sources, because 
they have electron scattering atmospheres needed for producing limb 
polarization through Thomson scattering \citep{Simmons95b}.
Unfortunately, the bulge of the Galaxy does not contain a large number of
hot giant stars.
However, polarization may be also induced by the scattering of star light off 
atoms, molecules and dust grains in the adsorptive atmospheres of evolved, 
cool stars as shown by \cite{Simmons02} and \cite{Ignace06}.
These more ubiquitous stars, that do not have levels of polarization as high 
as those predicted by the Chandrasekar model, may display an intrinsic 
polarization of up to several { percent}, 
due to the presence of stellar winds
that give rise to extended adsorptive envelopes. 
This is the case for many cool giant stars, in particular for the red giants. 
Such evolved stars constitute a significant fraction of the lensed sources 
towards the Galactic bulge, the LMC \citep{Alcock97} and the M31 galaxy
\citep{Sebastiano10}, making them valuable candidates for observing variable 
polarization during microlensing events.

Polarization measurements on ongoing microlensing events
can be useful for further characterizing them, 
for testing stellar atmosphere 
models  and to complement finite source measurements \citep{Gould94}. 
In this respect, the detection of a variable polarization leads to an 
independent measure of the angular Einstein radius $R_{\rm E}$ 
of the lens, the position 
angle of the lens and the velocity direction in the sky
\citep{Yoshida}.

Of course, since accurate polarization measurements cannot be obtained 
with a survey telescope, alert systems are necessary allowing other larger 
telescopes to take polarimetric measurements during a microlensing event.

For definiteness, we consider the observed sample of highly magnified, 
single-lens {\it transit} events 
\citep{Choi} and a subset of exoplanetary events observed towards the 
Galactic bulge \citep{Gaudi10}. 
For these events we calculate the polarization profiles as
a function of time taking into account the nature of the source stars.
As an illustration, we also consider the expected polarization signal
for the PA-99-N2 event towards M31 \citep{Paulin03}.

\section{Polarization Models}

\subsection{Basic equations}

Let us consider the linear polarization of light scattered in a stellar
atmosphere. Following the approach in \cite{Chandra60}
we define the intensities $I_l(\mu)$ and $I_r(\mu)$ emitted by the scattering 
atmosphere in the direction making an angle $\chi$ with the 
normal to the star surface and polarized as follows: 
$I_l(\mu)$ is the intensity in the plane containing 
the line of sight and the normal, 
$I_r(\mu)$ is the intensity in the direction perpendicular to this plane
(light propagates in the direction ${\bf r \times l}$). Here 
$\mu = \cos{\theta}$ and we are assuming that the  polarization is a 
function only of $\mu$. The center-to-limb 
variation in the polarization across the stellar disc originates 
from the contribution of the scattering opacity to the total opacity in the 
star atmosphere. The polarization typically increases from the center 
($\mu=1$) to the stellar limb ($\mu=0$) 
since we tend to see more scattered, hence polarized, light 
towards the limb. In the hot stars, electron (Thomson) scattering is 
 one cause of opacity. 
In the case of cool stars,  coherent (Rayleigh) scattering on atomic and 
molecular hydrogen and atomic helium, and scattering off dust grains provide 
the main source of the scattering opacity.

We choose a coordinate system in the lens plane 
with the origin at the center $(x_0,y_0)$ of 
the projected position of the source star. 
The $Oz$ axis is directed towards the observer, 
the $Ox$ axis is directed towards the lens (in single-lens events) and
oriented parallel to the star-to-planet separation (in binary events).
The location of a point $(x,y)$ on the star surface is determined by the 
angular distance $\rho$ from the origin of the coordinates and by the 
angle $\varphi$ with the $Ox$ axis
($x=x_0 + \rho \cos \varphi$ and $y=y_0 + \rho \sin \varphi$).
In the above coordinate system $\mu = \sqrt{1-\rho^2/R^2}$, 
where $R$ is the star angular size. 
Here and in the following all distances 
are given in units of the Einstein radius $R_{\rm E}$ of the total lens mass.

To calculate the polarization of a star with center at the position 
$(x_0,y_0)$ we 
integrate the unnormalized Stokes parameters and the flux over the star 
\citep{Simmons95a,Simmons95b,Agol96}
\begin{eqnarray}
F = F_0 \int_{0}^{2\pi} \int_0^R A(x,y) ~ 
I_+(\mu) ~\rho d\rho~ d \varphi~,
\label{flux} 
\end{eqnarray}
\begin{eqnarray}
F_Q = F_0 \int_{0}^{2\pi} \int_0^R A(x,y) ~ 
I_-(\mu)~\cos 2 \varphi ~\rho d\rho~ d \varphi~,
\label{fq} 
\end{eqnarray}
\begin{eqnarray}
F_U = F_0 \int_{0}^{2\pi} \int_0^R A(x,y) ~ 
I_-(\mu) ~ \sin 2 \varphi ~\rho d\rho~ d \varphi~,
\label{fu} 
\end{eqnarray}
where
$F_0$ is the unamplified star flux, $A(x,y)$ the point source 
amplification due to the lens system and
\begin{eqnarray}
I_+(\mu) = I_r(\mu)+I_l(\mu)~, 
\end{eqnarray}
\begin{eqnarray}
I_-(\mu)=I_r(\mu)-I_l(\mu) ~.
\end{eqnarray}
As usual, the polarization degree and the polarization angle
are \citep{Chandra60}   
\begin{eqnarray}
       P = \frac{ \sqrt{F_Q^2+F_U^2}} {F}~~~~,~~~~~~~~~~
\theta_P = \frac {1}{2} \tan^{-1} \frac {F_U}{F_Q}~~.
\end{eqnarray}
Clearly, neglecting other distorting factors,  
the stellar light will be unpolarized if we observe the source star 
without any gravitational lens effect since,
due to the symmetry of their expressions, the parameters 
$F_Q$ and $F_U$ are equal to zero.
The gravitational field produced by the lensing system breaks the
spherical symmetry over the projected stellar disc and thus results in a 
non-vanishing polarization.

For single-lens events the amplification is \citep{Einstein36,Pacz86} 
\begin{eqnarray}
A(x,y) = \frac{\rho_s+2}{\rho_s \sqrt{\rho_s^2+4}}~,
\label{A-1lente}
\end{eqnarray}
where 
\begin{eqnarray}
\rho_s^2 = \rho_0^2 + \rho^2 - 2 \rho_0 \rho \cos \varphi
\end{eqnarray}
is the angular distance between the considered surface element on the stellar 
disc and the lens position.   
Here $\rho_0$ ($\rho$) is the angular distance of the lens 
(the surface element) from the star center
\footnote{ In the case of microlensing by a single-lens system
the Stokes parameter $F_U$ vanishes in the adopted
system of coordinates defined
by the line joining the lens and the star, and 
the polarization position angle is always
perpendicular to that line.}.

In the case of binary events, we evaluate the amplification map 
$A(x,y)$ at any point in the source plane 
by using the Inverse Ray-Shooting method \citep{Kayser,Wambsganss}. 
In this case
the amplification map depends on the mass ratio  $q$ between the planet and the star and the star-to-planet separation $d$.

The other  quantities entering in the above equations are 
the coordinates of the source star center. These are given at any time $t$
in terms of the other lens system parameters:
the impact parameter $u_0$ (the minimum distance of the source star center
projected on the lens plane 
from the center of mass of the lens system), 
the maximum amplification time $t_0$, the Eintein time $t_{\rm E}$ and,
for binary events, 
the angle $\alpha$ that the source trajectory makes with the $Ox$ axis.   

To derive a quantitative expression 
of the functions $I_l(\mu)$ and $I_r(\mu)$ in the above 
equations one has to consider the relevant physical mechanism giving rise to 
the variable polarization degree.

\subsection{ The hot star case}

In the case of  polarization of stellar light 
induced by the  electron scattering in the
atmosphere of hot stars, the functions $I_l(\mu)$ and $I_r(\mu)$ have been 
evaluated { (assuming a plane-parallel atmosphere)}
by \cite{Chandra60} 
and their numerical values   
can be approximated 
\footnote{As noted by \cite{Bochkarev_83}, the precision of the
assumed approximations is better than $10^{-3}$.}
by the following expressions 
\citep{Bochkarev_83} 
\begin{eqnarray}
I_+(\mu) = \left( \frac{1+16.035 \mu +25.503 \mu^2} 
{1+12.561\mu+0.331\mu^2} \right)~,  
\label{pol7a0} 
\end{eqnarray}
\begin{eqnarray}
I_-(\mu) = \left(\frac{0.1171+3.3207 \mu +6.1522 \mu^2}
{1+31.416\mu+74.0112 \mu^2}\right) \left(1-\mu\right)~. \label{pol7a}
\end{eqnarray}

\subsection{ The late type main sequence star case}

The continuum spectrum of the  Sun is linearly polarized by coherent 
scattering on neutral hydrogen in its ground state (Rayleigh scattering) 
and, with a minor contribution,  by (Thomson) 
scattering on free electrons 
\footnote{A polarization in spectral lines is also
produced by coherent scattering in atomic bound-bound transitions.}. 
The polarization is maximum near the solar limb due to the most favorable 
geometry there.
\cite{Stenflo99} developed a theoretical study for the formation of 
the continuum polarization, identifying the relevant physical mechanisms and
clarifying their relative roles. The key physical quantities are  
the scattering coefficients and the temperature gradient in the stellar 
layer where polarization is formed. With a solar model atmosphere as input, 
the polarization is obtained by numerically
solving the transfer equation for polarized radiation. 
The results of their numerical calculations are 
approximated by semi-analytical formulas, as a function of $\mu$ and of 
wavelength $\lambda$  \citep{Stenflo05}
\begin{eqnarray}
\label{eqstenflo}
P(\mu) = q_{\lambda} \frac {(1-\mu^2)} {(\mu+m_{\lambda}) 
(I_{\lambda}(\mu)/I_{\lambda}(1))}~,
\label{pol_Stenflo}
\end{eqnarray}
 where $I_{\lambda}(\mu)/I_{\lambda}(1)$ represents the center-to-limb 
variation of the intensity. The dependence of 
$q_{\lambda}$ and $m_{\lambda}$ from the wavelength
is represented by the relations 
$log(q_{\lambda})= a_0 + a_1 \lambda + a_2 \lambda^2$
and $m_{\lambda}=b_0 +b_1 \lambda$, where the coefficients are 
determined by fitting solar observations for different
wavelengths and atmosphere models (see Table 1 in \cite{Stenflo05}).
In particular, in the I band $q_{\rm I}=4.2 \times 10^{-4} $ and 
$k_{\rm I} \equiv m_{\rm I}^{-1} =50$. 

Following \cite{Carciofi}, we adopt the model in eq. (\ref{eqstenflo}) for 
calculating the polarization profile { for late type stars} of any 
spectral type (F, G, K, M), for which polarization measurements do  not exist. 
For the total intensity we use the linear limb-darkening profile \citep{Choi}
\begin{eqnarray}
I_{\lambda}(\mu) =  
\left[ 1- \Gamma_{\lambda} \left(1-\frac{3}{2} \mu \right)\right]~,  
\label{linear_LD} 
\end{eqnarray}
where the parameter $\Gamma_{\lambda}$ 
depends on wavelength, spectral type, surface gravity 
and metallicity of the source star \citep{Claret}.

\subsection{The case for cool giant stars}

Polarization signals in microlensing of stars with extended 
circumstellar envelopes have been studied by \cite{Simmons02} 
for a single-lens and in binary lensing by \cite{Ignace06}. 
As emphasized in these works, the model is well suited to describe
polarization in evolved, cool stars that exhibit stellar winds  
significantly stronger than those of the Sun.
The scattering opacity responsible for producing the polarization is
the photon scattering on atomic and molecular species (Rayleigh) 
or on dust grains.

The scattering number density distribution in the stellar envelope was
parametrized by a simple power law 
\begin{eqnarray}
n(r) = n_h ~(R_h/r)^{\beta}~,
\end{eqnarray}
where $r = (\rho^2+z^2)^{1/2}$ is the radial distance
from the star center and $n_h$ is the number density at
the radius $R_h$ of a central cavity. Indeed,
since dust grains and molecules cannot form at temperatures higher 
than 3000 K $ -$ 4000 K, a circumstellar cavity was considered between the 
photosphere radius $R$ and the condensation radius $R_h$, where the 
temperature drops to 1500 K.

The star envelope was assumed optically thin and the relevant scattering 
cross-section was evaluated in the dipole approximation.
The relevant expressions for the Stokes parameters are given in the paper
by \cite{Simmons02}, where one can see that the polarization degree linearly 
depends on the optical depth $\tau_{\lambda}$. It has also been 
found that the polarization 
degree during microlensing can reaches values up to a few  percent,
provided that $\tau_{\lambda} \simeq 0.1$

An estimate of the the order of magnitude of $\tau_{\lambda}$ 
was also derived 
assuming that the gas density profile in the stellar wind
follows the law 
\begin{eqnarray}
\rho_{\rm gas}(r) = \frac{\dot{M}} {4~ \pi~ r^2~ v(r)}~,
\end{eqnarray}
with $v(r) \simeq v_{\infty} (1-R_h/r)^{\gamma}$ ($\gamma \simeq 1/2$),
and considering the main contribution to the gas opacity 
at temperature $T<10^3$ K.
If dust grains are the main source to the stellar opacity, 
the optical depth is \citep{Ignace08} 
\begin{eqnarray}
\tau_{\lambda}  = 2 \times 10^{-3} \sigma {\mathcal K}_{\lambda} 
\left(\frac{\dot{M}} { 10^{-9}~M_{\odot} {\rm yr}^{-1} }\right)
\left(\frac {30~km s^{-1}} {v_{\infty}}\right)
\left(\frac {24R_{\odot}} {R_h} \right)~,
\label{tau} 
\end{eqnarray}
where $\sigma\simeq 0.01$ is the dust-to-gas mass density ratio 
and ${\mathcal K}_{\lambda} \simeq 200$ cm$^2$ g$^{-1}$ is the dust opacity at 
$\lambda > 5500 \AA$  \citep{Ferguson}. 
This means that the wind mass-loss rate must be 
around $10^{-7}~M_{\odot}~{\rm yr}^{-1}$ to obtain 
$\tau_{\lambda} \simeq 0.1$.
This level of mass loss can be achieved only in AGB stars.
This class of stars constitutes the sources of only a few of the observed
microlensing events, while much more common are subgiants and red giant stars.
For a realistic estimate of the polarization level for these stars,
one has to relate $\tau_{\lambda}$ to the stellar parameters of the
magnified star. 
Indeed, it is well known that from main sequence to AGB phases, 
the mass-loss rate increases by 7 order of magnitude \citep{Ferguson}.
By performing simulations of the mass loss of intermediate and 
low-mass stars, it was  shown that the increase of mass loss is a
continuous function
(while the velocity and temperature evolutionary traces
show gaps between the subgiant and AGB phases) and that $\dot M$ obeys the 
relation \citep{Ferguson}
\begin{eqnarray}
\dot{M} = 2 \times 10^{-14}  
\frac{ (L/L_{\odot} ) ( R/R_{\odot} )^3 
( T/T_{\odot})^9 }  
{(M/M_{\odot})^2}~ M_{\odot}~{\rm yr}^{-1}~.
\label{dotM}
\end{eqnarray}
So, for the more common stars evolving
from main sequence to red giant star phases
(see also Fig. 15 in \cite{Suzuki}),
mass loss rates between $(10^{-13} - 10^{-8})~M_{\odot}~{\rm yr}^{-1}$ 
are expected. This corresponds to values of $\tau_{\lambda}$ in the range 
$4 \times 10^{-7} - 4 \times 10^{-2}$ and, therefore,
to polarization degree that are not always negligible, reaching values up 
to 0.5  percent. 

Here, we adopt the above formalism to derive the polarization profiles 
for source stars with envelopes.
To this aim, the Stokes parameters are numerically evaluated by using 
in eqs. (\ref{flux}), (\ref{fq}) and (\ref{fu}) the 
relevant expressions for $I_+(\mu,\varphi)$ and $I_-(\mu,\varphi)$,
as given by \cite{Ignace06}.

\section{Polarization results}

To make predictions of the polarization behavior during microlensing
we have generated a large sample of single and binary lens 
events. Assuming the nature of the source star 
{ (a hot star, a late type main sequence or a cool 
giant star)} and its radius $\rho_*$, 
we explore the multidimensional space for microlensing 
($u_0$, $t_E$, $R_E$) and planetary ($q$, $d$, $\alpha$) parameters
\citep{Ingrosso06,Ingrosso09}, 
with the aim of deriving the polarization signal and eventually its 
observability. 
\begin{figure}
\includegraphics[width=90mm]{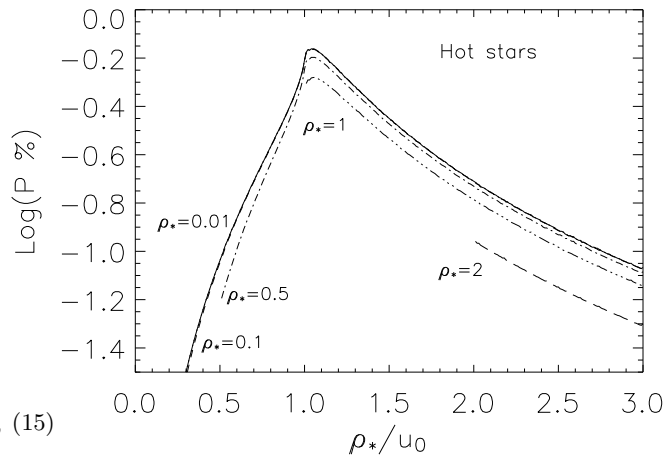}
\caption{Polarization degree $P$ in { percent} versus $\rho_*/u_0$, 
for different values of $\rho_* = 0.01,~ 0.1,~ 0.5,~ 1,~ 2 $ 
in the case of hot stars with polarization \`a la Chandrasekhar.}
\label{fig1} 
\end{figure}
\begin{figure}
\includegraphics[width=90mm]{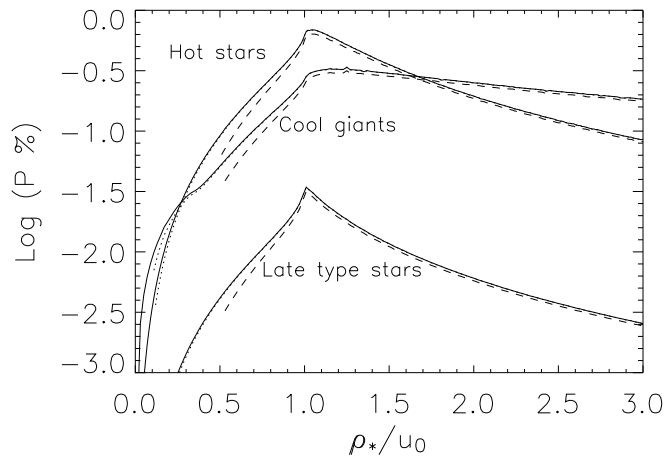}
\caption{Polarization $P$ in { percent} versus $\rho_*/u_0$ 
and $\rho_* = 0.01, ~0.1,~ 0.5$ for { hot stars, 
cool giants ($\tau=10^{-3}$, $\beta=2$, $R_h=5~R_*$) 
and late type stars} ($q_{\rm I}=4.2 \times 10^{-4} $, 
$k_{\rm I} =50$, $\Gamma_I =0.5$).} 
\label{fig2}
\end{figure}
For a given set of parameters of both the source and lens system,
we first evaluate the amplification map in the lens plane and then
the corresponding polarization curve, which shows a continuous variation 
with time, depending on the source position in the lens plane. 

For single-lens events, generally, the maximum polarization degree 
takes place at the time $t_0$ of the maximum amplification. { This happens 
when $\rho_*/u_0 < 1$ ({\it bypass} events) for which 
the source trajectory remains outside the lens. 
For single-lens events with $\rho_*/u_0 > 1$ ({\it transit} events) 
for which some part of the source is directly aligned
with the lens center and observer, the polarization curve has two maxima and 
one minimum, bracketed by the maxima, which coincides with the instant of 
maximum amplification \citep{Simmons95a,Simmons95b}.}  
In the case of {\it transit} events, the polarization signal gets the maximum 
value when the source disk enters and exits the fold caustic 
(two peaks appear at symmetrical position with respect to $t_0$).
In this case, the characteristic time scale $\Delta t_*$
between the two peaks of the 
polarization curve is related to the 
{\it transit} duration of the source disc
by \citep{Choi}
\begin{eqnarray}
\Delta t_* \simeq 2 t_E \times \sqrt{\rho_*^2-u_0^2}~.
\label{transit_time}
\end{eqnarray}

In binary microlensing events, 
for which there are a central (stellar) and
{ one or two planetary caustics depending on the star-to-planet separation
$d>1$ or $d<1$, respectively},
significant polarization 
may be induced when the star trajectory intersects one of the caustics.
Then, a polarization peak occurring  at $t \ne t_0$ implies that a 
planetary caustic is intersected by the source. 
Moreover, as for single-lens events, if the star source {\it bypasses} 
one caustic (central or planetary), a single polarization peak 
appears in the polarization profile at the time of nearest 
approach, while a double-peak feature occurs 
if the stellar disc {\it transits} a caustic.

Therefore, detection of polarization signals may allow,
in principle, to distinguish
between single and binary lens events.
Indeed, single and binary events
can be separated by the presence of a polarization signal at $t \ne t_0$ or
by an eventual asymmetry in the polarization curve (near $t_0$)
induced by the planetary caustic. This happens for events 
with $b \simeq 1$, for which the planetary caustic shifts in a position
close to the central one. 

{ In general, the polarization signal in binary microlensing events
depends on the projected source trajectory in the lens plane, the 
size and position
of the caustics and the diameter of the source disc. In Section 3.2
we consider, as an example, some exoplanetary events well observed
towards the Galactic center, leaving a detailed study of the
polarization in binary microlensing events to a subsequent analysis.}

For single-lens events and polarization induced by electron scattering 
in the atmosphere of hot stars,
in Fig. \ref{fig1} we show the polarization degree $P$ (in { percent})
as a function of the ratio 
$\rho_*/u_0$. 
As one can see, for $\rho_*/u_0  \ut > 0.3$ and 
$\rho_* \ut < 0.1$, the polarization degree depends 
only on the ratio $\rho_*/u_0$ (see also Fig. 3 in \cite{Yoshida}). 
The polarization degree has the maximum value 
$P_{\rm max} \simeq 0.7$ { percent}  at $\rho_*/u_0\simeq 1.04$ and  
for low $\rho_*$ values. 
For large values of $\rho_*$  the polarization degree 
decreases as the star radius increases.
We note that, similarly to limb-darkening measurements, 
the polarization degree is maximized 
for events with large finite source effects, namely for $\rho_*/u_0 \simeq 1$.
\begin{figure}
\includegraphics[width=90mm]{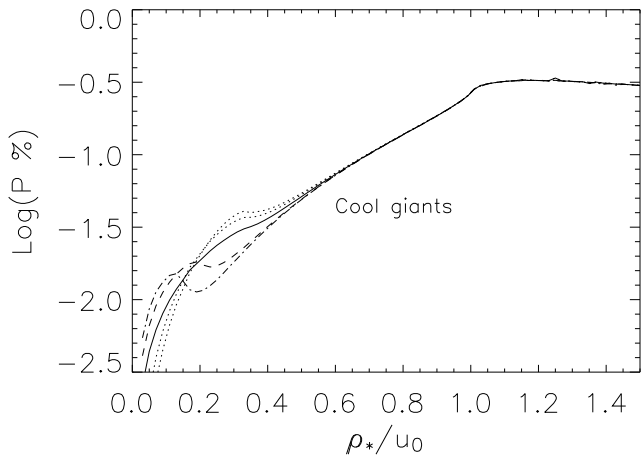}
\caption{Polarization versus $\rho_*/u_0$ in the case of 
cool giant stars. Assuming $\tau=10^{-3}$, we show the 
effect of varying the model parameters. 
{ As a reference, the solid line corresponds to $R_h=3~R_*$ and $\beta=2$. 
Dashed lines are for models with $\beta=2$ and
increasing values of $R_h=5~R_*$ and $R_h=7~R_*$. For these models
a secondary polarization peak occurs at sequentially smaller 
values of $\rho_*/u_0 <1 $, as $R_h$ increases.
Dotted lines are for models with $R_h=3~R_*$ and increasing
values of $\beta=3$ (bottom) and $\beta=4$ (upper). }}
\label{fig3}
\end{figure}
\begin{figure}
\includegraphics[width=90mm]{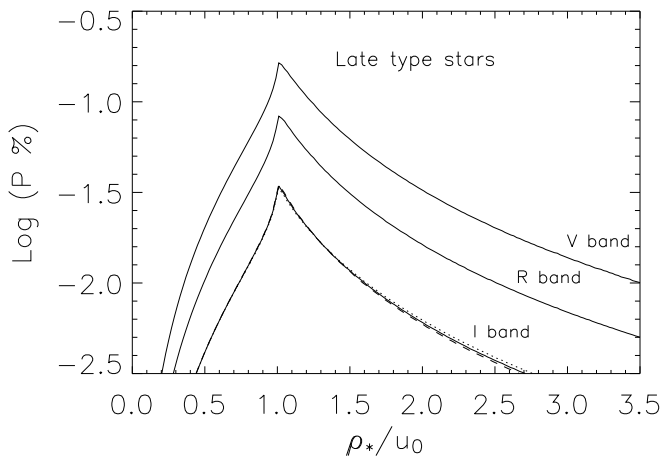}
\caption{Polarization versus $\rho_*/u_0$ { for late type stars}.
Assuming $\rho_*=0.01$ and $\Gamma=0.5$, we show
the polarization versus $\rho_*/u_0$ in the V, R, I band (solid lines).
For the I band case the polarization is also shown 
for $\Gamma = 0.1$ (dotted line) and 0.9 (dashed line).}
\label{fig4} 
\end{figure}

To compare the polarization degree for the three considered
source models { (hot, late type main sequence and cool giant stars)}, in 
Fig. \ref{fig2} we show $P$ versus $\rho_*/u_0$.
As in Fig. \ref{fig1},  the maximum polarization occurs 
at $\rho_*/u_0 \simeq 1$ and for low values of $\rho_*$. 
The polarization { for late type stars} is calculated 
in the I-band assuming $q_{\rm I}=4.2 \times 10^{-4} $, 
$k_{\rm I} =50$ and  $\Gamma_{\rm I}=0.5$, 
that are typical parameter values for Sun like stars.
As one can see the maximum polarization degree is in any case much lower
than that for hot star sources. 
In the case of cool giant stars the polarization profile
is not much different with respect to that for hot stars if  
$\tau \simeq 10^{-3}$. Of course, even larger values for $P$ can be obtained, 
provided that $\tau > 10^{-3}$.
In Fig. \ref{fig2} we also see that there exists a limiting value
of the ratio $\rho_*/u_0$ ($\simeq  1.7$ for $\tau = 10^{-3}$)
at which the polarization degree for cool giant stars
overcomes that for hot stars. This is a relevant point since,
as we shall see later on, so large values of $\rho_*/u_0$
are rather common in the microlensing events 
observed towards the Galactic bulge.

In Fig. \ref{fig3} for cool giant stars 
we show how the polarization degree depends on the relevant
model parameters 
{ \citep{Simmons95a,Simmons02}}. 
{ We take $\rho_*=0.01$ and  $\tau=10^{-3}$ and let 
the circumstellar envelope  parameters $\beta$ and $R_h$ vary. 
For $\rho_*/u_0 > 0.5$ the polarization degree is
almost insensitive to the $\beta$ and $R_h$ values, 
since the polarization is mainly due to the
source star disc}.

In Fig. \ref{fig4}, for { late type stars}, we show how the polarization 
degree depends on the ratio $\rho_*/u_0$ and the parameters 
$\Gamma_{\lambda}$, $q_{\lambda}$ and $k_{\lambda} \equiv m_{\lambda}^{-1}$
in eqs. (\ref{pol_Stenflo}) and (\ref{linear_LD}). To this aim, 
we consider the V, R, I bands for which, adopting the relations 
and the parameter values in \cite{Stenflo05}, we obtain 
           $q_V=2.5 \times 10^{-3}$ , $k_V =18$ (in the V band),
           $q_R=1.2 \times 10^{-3}$ , $k_R =23$ (in the R band),
           $q_I=4.2 \times 10^{-4}$ , $k_I =50$ (in the I band).
As one can see in Fig. \ref{fig4} the polarization in  { late type stars}
is almost insensitive to variation of $\Gamma$, 
while it rapidly decreases from the V to I band.
In this respect, polarization observations towards  { late type stars}
should be performed at low wavelength, at least for near enough objects 
for which absorption by interstellar medium is negligible. However, 
for polarization measurements towards Galactic bulge stars, the I band
represents the best { compromise} between interstellar absorption and 
the dependence of the polarization on wavelength. 

Concerning the polarization in exoplanetary events, 
as already mentioned, we find that one or more spikes appear 
in the polarization profiles when the caustics are crossed by the source star 
disc, corresponding to a large amplification of a narrow piece 
of the star surface. Outside
the caustics, the polarization signal from the total stellar flux is 
almost vanishing.  Clearly, in a real observation,  the obtained polarization
profile, in addition to the usual light-curve observations,
may be used to constrain the binary system parameters.

\begin{table*}
\centering
\caption{Parameters of 11 highly magnified single-lens events towards the 
Galactic bulge with identified source star type.}
\medskip
\begin{tabular}{c|c|c|c|c|c}
\hline
\hline
event            & source &  $u_0    $ & $t_E$ & $\rho_*$                           & Reference \\
                 & type   & ($10^{-3}$)& (day) & ($10^{-3}$)                        &           \\
\hline
OGLE-2007-BLG-224/MOA-2007-BLG-163 & F V             &0.29& 6.91& 0.9               &  Gould et al. (2009)      \\
OGLE-2008-BLG-279/MOA-2008-BLG-225 & G V             &0.66&106.0& 0.68              &  Yee  et al. (2009)  \\
MOA-2009-BLG-174                   & F V             &0.5 &64.99& 2.0               &  Choi et al. (2012)  \\
MOA-2011-BLG-274                   & G V             &2.8 & 2.78& 12.3              &  Choi et al. (2012)  \\
\hline                     
\hline                     
OGLE-2004-BLG-254                  & K III           &6.1 &12.89& 40.6              &  Cassan et al. (2006)/Choi et al. (2012)\\
OGLE-2004-BLG-482                  & M III           &1   & 9.61&13.09              &  Zub et al. (2011)  \\
MOA-2007-BLG-176                   & K III           &36  & 8.09&59.4               &  Choi et al. (2012)  \\
OGLE-2007-BLG-302/MOA-2007-BLG-233 & G III  & 5.4 & 15.92  & 36.8                   &  Choi et al. (2012)  \\
OGLE-2008-BLG-290/MOA-2008-BLG-241 & K III                  &2.76  & 16.36  & 22.0  &  Choi et al. (2012)  \\
MOA-2011-BLG-093                   & G III                  &28.8  & 14.97  & 53.3  &  Choi et al. (2012)  \\
OGLE-2011-BLG-1101/MOA-2011-BLG-325& K III                  &47.4  & 29.55  & 96.0  &  Choi et al. (2012)  \\
\hline
\hline
\end{tabular}
\label{table1}
\end{table*}
\begin{table*}
\centering
\caption{Calculated parameters for the same events in Table \ref{table1}. 
Here $I_0$ is the I band magnitude at $t_0$ and 
$P_{\rm max}$ is the maximum value of the polarization.}
\medskip
\begin{tabular}{c|c|c|c|c|c|c}
\hline
\hline
event            & $I_0$  & $\Delta t_*$ & $\rho_*/u_0$ & $\Gamma_{\rm I}$ &  $\tau_{\rm I}$  & $P_{\rm max}$  \\
                 & (mag)  & (day)      &               &                  &  $(10^{-4})$           & (\%)           \\
\hline
OGLE-2007-BLG-224/MOA-2007-BLG-163 &10.4&0.01       &  3.10 & 0.44 & - & 0.02\\
OGLE-2008-BLG-279/MOA-2008-BLG-225 &12.6&0.03       &  1.03 & 0.52 & - & 0.04\\
MOA-2009-BLG-174                   &11.9&0.25       &  4    & 0.33 & - & 0.04\\
MOA-2011-BLG-274                   &13.3&0.07       &  4.39 & 0.52 & - & 0.03\\
\hline
\hline
OGLE-2004-BLG-254                  & 12.0   & 1.03  &   6.65        &-            & 7 & 0.23 \\
OGLE-2004-BLG-482                  & 11.4   & 0.25  &  13.09            &-        & 6 & 0.20 \\
MOA-2007-BLG-176                   & 14.4   & 0.76  &   1.65       &-             &100& 3.23 \\
OGLE-2007-BLG-302/MOA-2007-BLG-233 & 11.9   & 1.16  &   6.81      &-              & 35& 1.14 \\
OGLE-2008-BLG-290/MOA-2008-BLG-241 & 9.7    & 0.71  &   7.97            &-        & 2 & 0.07 \\
MOA-2011-BLG-093                   & 12.2   & 1.34  &   1.85      &-              & 5 & 0.16 \\
OGLE-2011-BLG-1101/MOA-2011-BLG-325& 11.9   & 4.93  &   2.03       &-             & 20& 0.65 \\
\hline
\hline
\end{tabular}
\label{table2}
\end{table*}

To be more explicit on the expected polarization variability in microlensing, 
we now focus on a sample of highly magnified {\it transit} events 
(both single-lens and exoplanetary) observed towards the Galactic bulge 
by the OGLE and MOA collaborations \citep{Choi}. 
We also consider some {\it bypass } 
exoplanetary events towards the Galactic bulge
for which the polarization signal is enhanced by the source star 
transiting a planetary caustics.
The parameters of the best fit models to the light-curves are given
for single-lens events in Table \ref{table1} and 
for exoplanetary events in Table \ref{table3}.

\subsection{Single-lens events towards the Galactic center}
 
In Table \ref{table1} we give the parameters of 11 highly
magnified single-lens events, selected among the 17 given in the Table 3  
in \cite{Choi}.
The selected events are characterized by a clearly identified star source
for which the physical parameters have been measured, also by follow up
observations. Among them, 4 source stars are  { late type stars} and  
7 are cool giant stars.
All events are characterized by very small values of 
their impact parameter ($u_0 \ll 1$) and relatively large source size,
so that $\rho_*/u_0>1$. In these events the lens crosses over the
source star disc  and therefore, at the peak of the event, different parts 
of the source are magnified by substantially different amounts. 
This fact provides a rare chance not only to measure the 
brightness profile of a remote star, as shown by \cite{Choi}, but also to 
maximize the polarization effects.

\subsubsection{ Late type main sequence source stars}

Let us start by considering a subsample of the 4 events in 
Table \ref{table1} that consists of 
sources belonging to the 
population of { late type main sequence stars} (F and G spectral type).
We evaluate the polarization profiles as a function of time
by using in eqs. (\ref{flux}), (\ref{fq}) and (\ref{fu}) the 
polarization law $P(\mu)$ in eq. (\ref{pol_Stenflo}),
assuming the solar value for the parameters
$q_{\rm I} = 4.2 \times 10^{-4} $ and  
$k_{\rm I} = 50$.
The adopted limb-darkening profile is given in eq. (\ref{linear_LD})
where the values of the $\Gamma_{\rm I}$ coefficients 
are evaluated in the I band and given in Table \ref{table1}.
The polarization profiles of these 4 events, that are shown in Fig. \ref{fig5},
 have a unique behavior characterized by the presence of a double peak that 
corresponds to the time instants at which the lens  {\it enters} and 
{\it exits} the source star disc.
The time interval between the two peaks, as calculated by
eq. (\ref{transit_time}), is given in Table \ref{table2},
where we also give the I band magnitude $I_0$ at the instant $t_0$ of maximum
magnification. We note that, due to the typical small values of 
$\Delta t_* \ut < 1$ day, the source magnitude at the instant of maximum 
polarization is always very close to $I_0$. 
The maximum polarization value (occurring in correspondence
of each peak) turns out to be  $\simeq 3 \times 10^{-2}$ { percent},
{ almost irrespective} of the $u_0$ and $\rho_*$ values,
since the ratio $\rho_*/u_0>1$. 
Indeed, the ratio $\rho_*/u(t)$ is a function of time and 
always exists a time value $t_{\rm max}$ at which
$\rho_*/u(t_{\rm max}) = 1.04$ where, as already shown, 
the polarization degree gets the maximum value.
In between the two peaks the polarization profile depends on the ratio 
$\rho_*/u_0$ (giving the extent of finite source effects) and 
rapidly flattens as $\rho_*/u_0$ increases.

\subsubsection{Cool giant source stars}

\begin{figure}
\includegraphics[width=90mm]{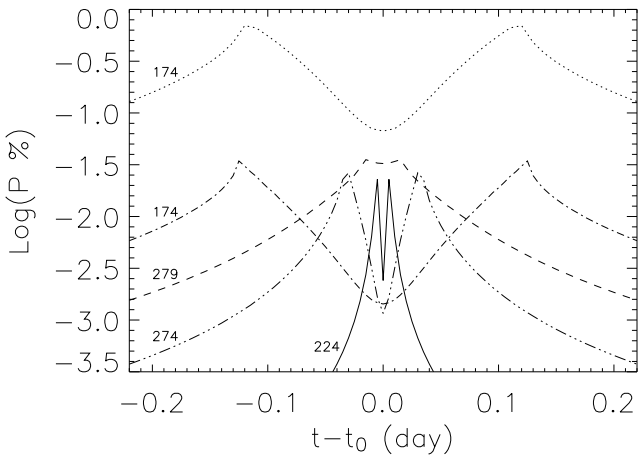}
\caption{Polarization profiles for highly magnified single-lens
events, with { late type source stars}. 
{ For comparison purpose, assuming the same microlensing
parameters of the MOA-2009-BLG-174 event,
we show the polarization profile (dotted line) that one could expect 
in the case of a hot source star 
giving rise to a polarization \`a la Chandrasekhar.}}
\label{fig5}
\end{figure}

\begin{figure}
\includegraphics[width=90mm]{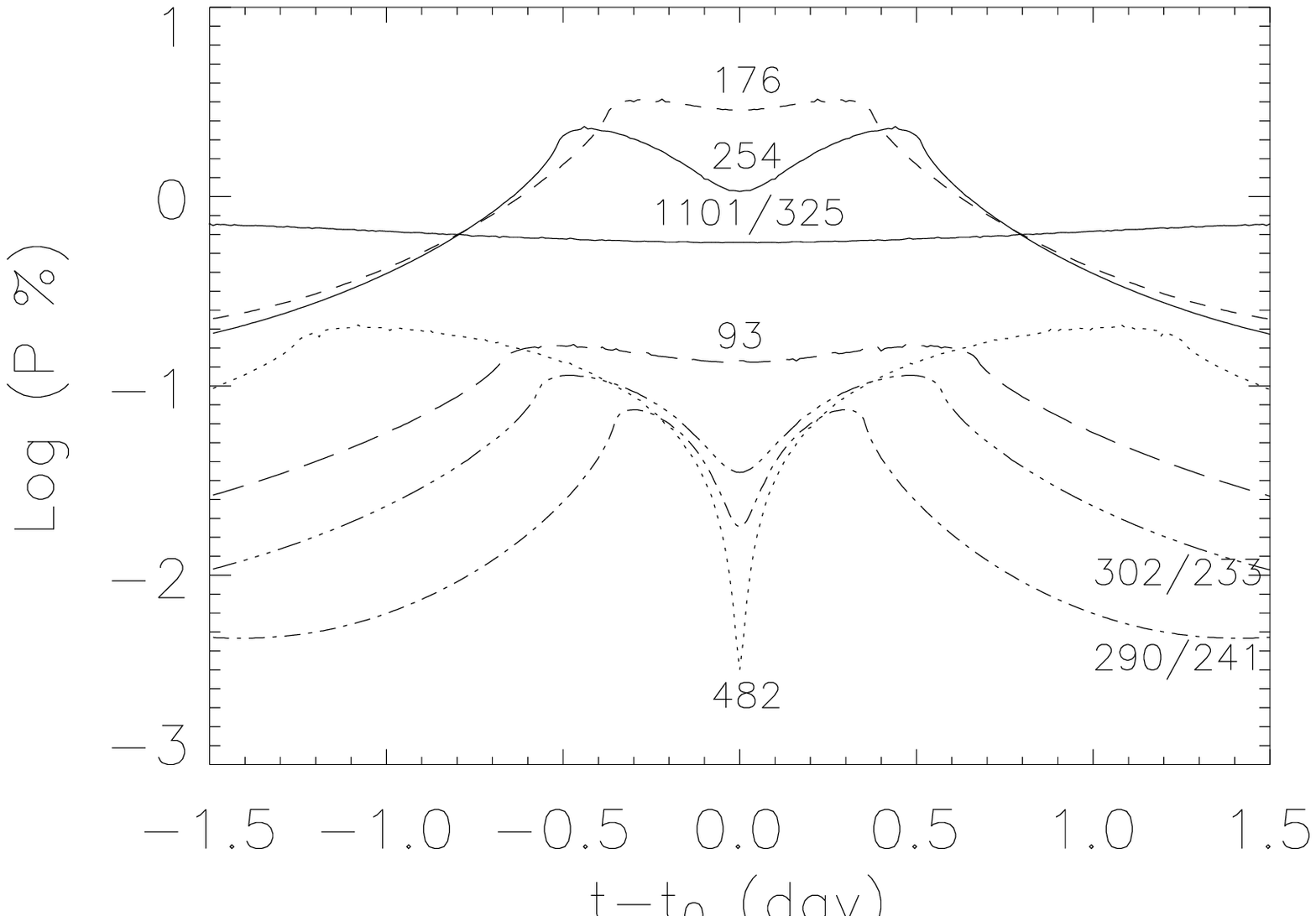}
\caption{Polarization profiles for the 7 single lens events
with giant source stars given in Table \ref{table2}.
For the event OGLE-2011-BLG-1101/MOA-2011-BLG-325
the maximum value of polarization ($P_{\rm max}=0.65$ { percent})
occurs at $t-t_0 \simeq \pm 2.5 $ days.}
\label{fig6}
\end{figure}

For the seven single-lens events in Table \ref{table1} 
whose source is a cool giant star, 
we calculate the polarization profile 
following the formalism of Subsection 2.4. 
For simplicity we fix the parameter values $\beta=2$ and $R_h=5R_*$, 
{ irrespective} of the source spectral type
\footnote{As shown in Fig. \ref{fig3}, for $\rho_*/u > 0.5$
and $\rho_* < 0.01$, the polarization degree is almost independent on the 
adopted parameter values, while it slightly depends on them otherwise.}.
Of course, in this case the polarization degree depends on the
optical depth $\tau_{\rm I}$ that is estimated from the $\dot M$ value
in eq. (\ref{dotM}). It turns out that our sources are characterized
by a stellar envelope with $\tau_{\rm I}$ in the range
$10^{-4} - 10^{-2}$. The corresponding polarization curves 
are shown in Fig. \ref{fig6}. 
As for single-lens events with  { late type main sequence source stars} 
(see Fig. \ref{fig5}), two polarization peaks are present (occurring 
when the lens enters and 
exits the source star disc) and the duration of the signal
is related to the lensing 
parameters by eq. (\ref{transit_time}). 
We note in particular that the flat polarization profile
in the case of the event 
OGLE-2011-BLG-1101/MOA-2011-BLG-325 is due to the large value of
$\Delta t_* \simeq 5$ day, which is larger of the time interval 
(3 days around $t_0$) considered in the figure.
In Table \ref{table2} we give for all the events
the values of the magnitude $I_0$ at the 
peak of the microlensing event, the time duration $\Delta t_*$, 
the optical depth $\tau_{\rm I}$ in the I band
and the maximum value of polarization $P_{\rm max}$.

\subsection{Exoplanetary events towards the Galactic center}

Planetary perturbations on the single-lens magnification 
pattern are expected to fall in two classes.  
The first class of perturbations occurs when a source star passes 
near a planetary caustic.
Although these perturbations are more common, they are also 
unpredictable  and can occur at any time during the event.
The second class of planetary perturbations occurs in highly 
magnified events in which the source becomes very closely aligned with 
the primary central caustic. 
These events allow to probe the region of the central
caustic (associated to the primary lens), that may be perturbed even by 
low-mass planets lying anywhere sufficiently close to the Einstein ring.
Since in this case the time of occurrence of the event  peak can 
be predicted in advance, it is possible  to focus  
observing resources to make intensive observations around the peak
of the event.
This is, indeed, the strategy that has been adopted 
by the $\mu$-FUN collaboration \citep{microfun} and
that has allowed the discovery
of half of the exoplanets detected by microlensing \citep{Gaudi10}. 
It goes without saying that this class of events may provide 
the highest chance to
detect the polarization signals,
at least for events showing large finite source
effects, namely with $\rho_*/u_0 >1$.

\subsubsection{High magnification events}

\begin{table*}
\centering
\caption{Parameters of 6 exoplanetary events towards the Galactic bulge.}
\medskip
\begin{tabular}{c||c|c|c|c|c|c|c|c|}
\hline
\hline
event                   & source &  $u_0$    &  $q$       &  $d$     & $t_E$ &$\alpha$& $\rho_*$  & Reference \\
                        & type   &$(10^{-3})$& $(10^{-3})$&          & (day) & (deg)  &$(10^{-3})$& \\
\hline
\hline
MOA-2007-BLG-400  &LT/subgiant&  0.25 & 2.6   &  0.34    & 14.41 & 227.06           & 3.26      & Dong et al., 2009\\
                        &        &  0.27 & 2.5   &  2.87    & 14.43 & 226.99          & 3.29  &                   \\
MOA-2008-BLG-310  &G V  &  3    & 0.331 &  1.085   & 11.14 & 69.33               & 4.93 & Janczak et al., 2010\\
                        &        &  3.01 & 0.32  &  0.927   & 11.08 & 69.33           & 4.95  & \\
OGLE-2005-BLG-169 &G V  &  1.24 & 0.086 &  1.0198  & 42.27 & 117.00          & 0.44 & Gould et al., 2006\\ 
                        &        &  1.25 & 0.083 &  0.9819  & 42.09 & 122.65          & 0.39  & \\
\hline
\hline
OGLE-2005-BLG-390   & K III&  35.9 & 0.076 &  1.610   & 11.03 & 157.91            & 25.6      &Beaulieu et al., 2006\\ 
OGLE-2003-BLG-235/MOA-BLG-53 
         & G IV     &  133  & 3.9   &  1.120   & 61.5  & 223.8         & 0.96      & Bond et al., 2004 \\
MOA-2007-BLG-368  & G V   &  82.5 & 0.127 &  0.9227  & 53.2  & 25.90            & 1.88      & Sumi et al., 2010 \\
\hline
\end{tabular}
\label{table3}
\end{table*}
\begin{table*}
\centering
\caption{Calculated parameters for the events in Table \ref{table3}.}
\medskip
\begin{tabular}{c|c|c|c|c|c|c|}
\hline
\hline
event                   &$I_0$  & $\Delta t_*$ & $\rho_*/u_0$ & $\Gamma_{\rm I}$& $\tau$     &  $P_{\rm max}$    \\
                        &(mag)  & (day) &                        &                   & $(10^{-4})$&  (\%)             \\
\hline
\hline
MOA-2007-BLG-400                & 9.9   & 0.10   & 13.04                           &0.47 & -       & 0.02      \\
MOA-2008-BLG-310                & 12.9  & 0.09   & 1.64                            &0.52 & -       & 0.03     \\
OGLE-2005-BLG-169               & 13.1  & -      & 0.35                            &0.52 & -       & 0.002      \\ 
\hline
\hline
OGLE-2005-BLG-390               & 13.1   & -      & 0.71                           &-    & $10^{-3}$&0.10  \\ 
OGLE-2003-BLG-235/MOA-BLG-53    & 17.6   & -      & 0.007                          &0.52 & -        &0.01   \\
MOA-2007-BLG-368                & 15.4   & -      & 0.02                           &0.52 & -        &0.02  \\
\hline
\end{tabular}
\label{table4}
\end{table*}

\begin{figure}
\includegraphics[width=90mm]{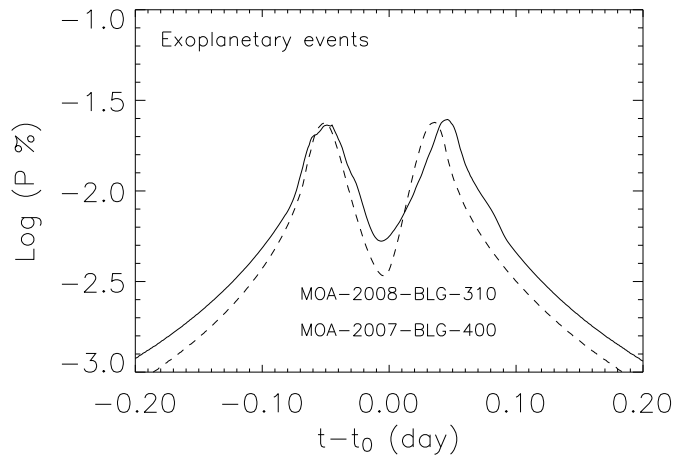}
\caption{Polarization profiles for the exoplanetary events
MOA-2008-BLG-310 (solid line) and MOA-2007-BLG-400 (dashed line). 
In both cases the lens system {\rm transits} the source star
and the polarization is calculated for the  { late type source star} 
case.}
\label{fig7}
\end{figure}

\begin{figure}
\includegraphics[width=90mm]{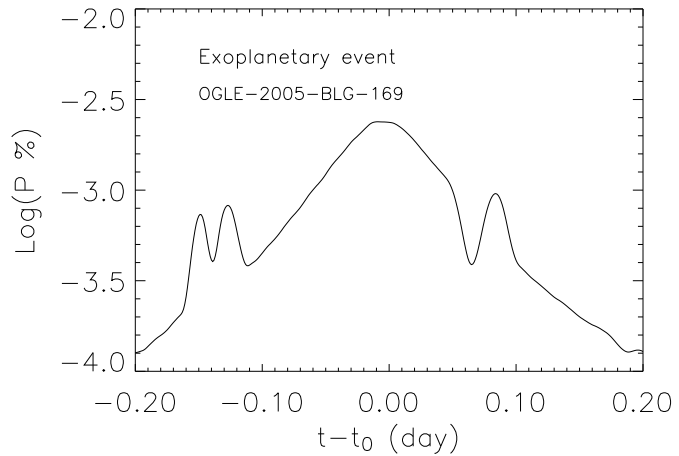}
\caption{Polarization profile for the exoplanetary event
OGLE-2005-BLG-169. The lens {\rm bypasses } the source star.}  
\label{fig8}
\end{figure}

{ Up to now} 15 exoplanets have been discovered by microlensing towards
the Galactic bulge \footnote{See http://exoplanet.eu.}
in 14 systems. Among  them, we consider in Table 
\ref{table3} six events
representative of the two mentioned classes of planetary perturbations.
Let us start by considering the first 3 highly magnified exoplanetary events 
belonging to the first class, 
for which the source star passes near the primary 
lens. The events MOA-2007-BLG-400 \citep{Dong} and
MOA-2008-BLG-310 \citep{Janczak10} are {\it transit} events, while
OGLE-2005-BLG-169 \citep{Gould06} is a {\it bypass} event.
These events are characterized by  { late type (LT)  source stars},
even if the first source  might be a subgiant star
close to the turn off point.
We calculate the expected polarization signal following Subsection 2.3,
assuming  $q_{\rm I}=4.2 \times 10^{-4} $ and  
$k_{\rm I} =50$ as for the  Sun.
The corresponding polarization profiles are shown in 
Fig. \ref{fig7} (MOA-2007-BLG-400 and MOA-2008-BLG-310)
 and Fig. \ref{fig8} (OGLE-2005-BLG-169).

As one can see, the polarization curves in  
Fig. \ref{fig7} appear similar to those 
in Fig. \ref{fig5} for single-lens events 
with source stars of the same nature, even if 
the presence of the planet in this case
produces a little asymmetry with respect to $t=t_0$, that is not present
in Fig. \ref{fig5}. This might offer in principle an independent
test of the exoplanet presence from polarization measurements.

The third event 
in Table \ref{table3} is a highly magnified
{\rm bypass} event with a Neptune mass ratio $q\simeq 8.6 \times 10^{-5}$ 
planetary companion to the primary lens star \citep{Gould06}. This low mass 
planet is lying sufficiently close to the Einstein ring and so it is able to 
perturb the light-curve expected for a single-lens event.  
As one can see in Fig. \ref{fig8}, 
the polarization profile shows { the presence of secondary peaks occurring}
when the star disc enters and exits the small central caustic associated
with the planet and closely aligned with the host star.
The features in the polarization curve closely resemble the residual
pattern of the light-curve (see Fig. 1 in \cite{Gould06}) and could be used
in principle to independently constrain the binary system parameters.
We would like to note that for this specific event the low value of the ratio 
$\rho_*/u_0=0.38$ determines a rather low value of the polarization degree.
{ An  exoplanetary event} with the same microlensing parameters but 
with larger finite size effects, should { give a higher}
polarization degree potentially detectable.

\subsubsection{Low magnification events}

The last 3 events in the Table \ref{table3} 
are exoplanetary events of low magnification, in which the source star
disc crosses a planetary caustic located far from host star, 
causing a peak in the polarization profile at $t \ne t_0$. 
The polarization is evaluated 
following the formalism in Section 2.3 and 2.4, depending on 
the nature of the source star.

\begin{figure}
\includegraphics[width=90mm]{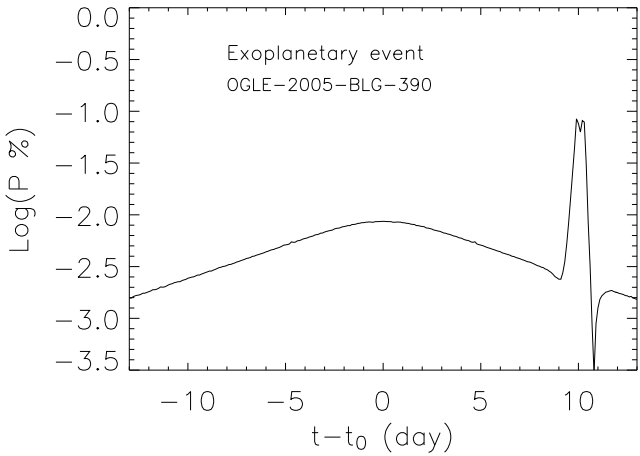}
\caption{Polarization profile for the exoplanetary event
OGLE-2005-BLG-390. } 
\label{fig9}
\end{figure}

In Fig. \ref{fig9} for the event
OGLE-2005-BLG-390 with a K giant source star, 
we use a value of $\tau \simeq 10^{-3}$ which results from 
eqs. (\ref{tau}) and (\ref{dotM}),
once the measured source star parameter values \citep{Beaulieu06} are used. 
{ As one can see, the sharp peak in the polarization profile
occurs at the same time ($t\simeq 10$ day)
of the planetary perturbation present in the light-curve 
(Fig. 1 in \cite{Beaulieu06}). 
However, the relative increase of the polarization signal
(with respect to the expectation for the single-lens model) 
is by far larger than the corresponding one in the light curve.
We also note that the polarization peak shows 
a two subpeak structure caused by the source star disc
transiting the planetary caustic. 
The size of this caustic, which turns out to be 
$ \ut < \rho_*$, may be computed  
by using the eqs. (8) and (9) given by \cite{Han06}.
The sharp depolarization following the polarization peak
in Fig. \ref{fig9} is due to a deamplification region
(following the planetary caustic) intersected by the source star disc.}

\begin{figure}
\includegraphics[width=90mm]{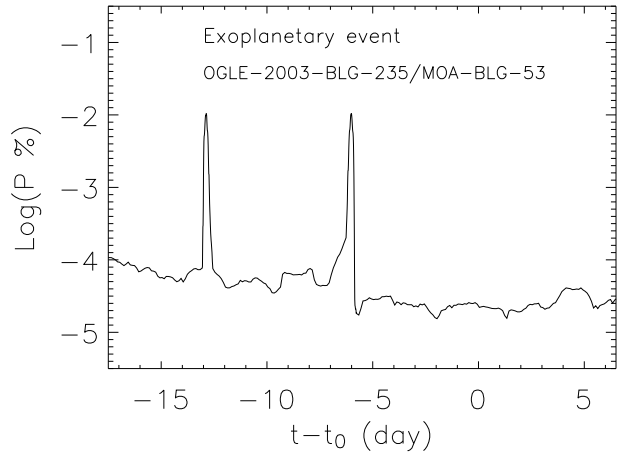}
\caption{Polarization profile for the exoplanetary event
OGLE-2003-BLG-235/MOA-BLG-53.} 
\label{fig10}
\end{figure}

The polarization profile evaluated for the exoplanetary event
OGLE-2003-BLG-235/MOA-BLG-53 \citep{Bond04}
is shown in Fig. \ref{fig10}, where we follow the approach described in 
Subsection 2.3 for { late type main sequence stars.}
{ As one can see the polarization degree is almost vanishing, except
for two sharp peaks, also present in the light-curve  
(see Fig. 1 in \cite{Bond04}),
occurring when the source passes over two folds 
of the planetary caustic.
The size of this caustic is much larger
than the projected source star disk, which  
enters the caustic at $t \simeq -13$ day and
exits at $t \simeq -6$ day.}

\begin{figure}
\includegraphics[width=90mm]{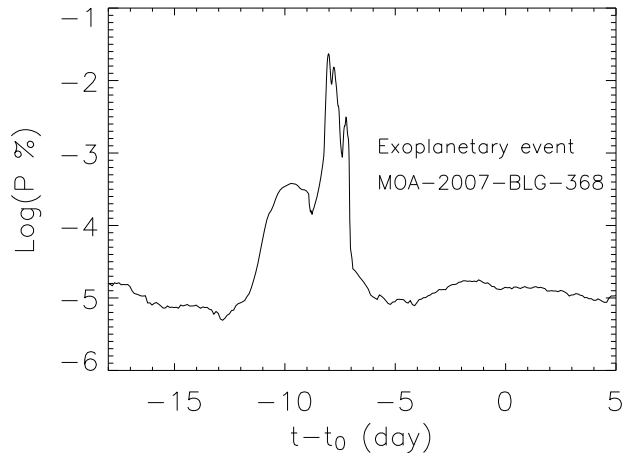}
\caption{Polarization profile for the exoplanetary event
MOA-2007-BLG-368.} 
\label{fig11}
\end{figure}

A similar behavior  occurs for the last considered 
exoplanetary event MOA-2007-BLG-368 \citep{Sumi10}
as shown in Fig. \ref{fig11}.


\subsection{Polarization profile for the PA-99-N2 event towards M31}
\begin{figure}
\includegraphics[width=90mm]{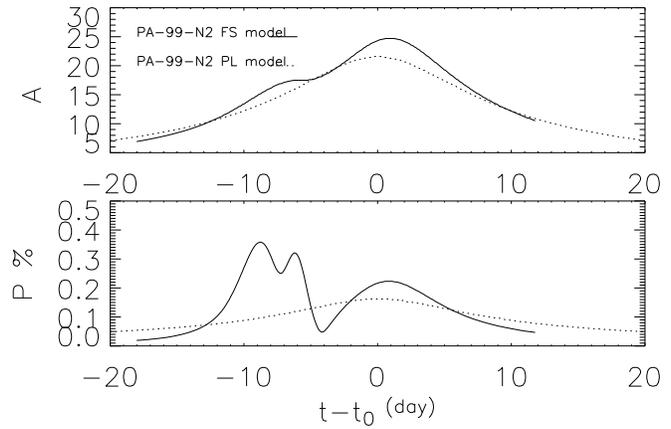}
\caption{Magnification (upper panel)  and polarization profiles (lower panel) 
for the event PA-99-N2 towards M31.
Solid (dotted) lines are used for the FS (PL) model in Table 2 in An et al.
(2004).}
\label{fig12}
\end{figure}

The PA-99-N2 event is a pixel lensing event observed in 1999 towards M31
by the Point-Agape Collaboration \citep{Paulin03}. The light curve 
shows a deviation from the Paczy\'nski law
which has been thoroughly analyzed by \cite{An04}.
In particular, \cite{An04} concluded in favor of a binary 
lens system with the primary lens companion being 
in the substellar mass range.
A further Monte Carlo study \citep{Ingrosso09,Ingrosso10} has
evidenced that one of the binary component may even  be an exoplanet
of a few Jupiter masses.
Although the analysis of \cite{An04} does not provide a strong evidence
for finite size source effect, which is necessary to the enhancement
of the polarization signal, to our purposes we consider
the lensing parameter for the { binary finite source (FS) solution}
provided in Table 2 { in the paper by \cite{An04}}. 
In particular, for the source 
we consider a red giant with $R\simeq 85~R_{\odot}$
and $T_{\rm eff} \simeq 3700$ K.
Accordingly, from eqs. (\ref{tau}) and (\ref{dotM}) 
we estimate $\tau \simeq 10^{-2}$ and using the formalism in Subsection 2.4 
we { compute} 
\footnote{ We note that the polarization $P$ has been always calculated
by considering only the flux $F$ from the lensed source star and 
neglecting any contribution of non-lensed sources (blending).
In this respect, our polarization results for the PA-99-N2 event 
have to be considered as upper limits,
since a strong blending component is expected
in pixel-lensing observations towards M31.}
the corresponding polarization profile shown 
in Fig. \ref{fig12}.
Inspection of this figure shows that the amplitude
of the polarization degree can reach values up to 0.5 { percent} 
(when the caustic region is crossed by the source star).
In the same figure, for comparison, we show the polarization curve 
(dotted line) expected for a 
{ point lens (PL) model}
with the same finite source effect.

Of course, with a distance modulus to M31 larger by about 10 mag
than that of Galactic bulge stars, the observability of the polarization 
for M31 sources remains, at the moment, highly speculative.

\section{Discussion and Conclusions }

The combined effects of large magnification and finite size source effects
in some microlensing events may allow to get relatively large values of the
polarization of the light from the source stars. In this paper
we calculate the polarization profile as a function of the time for a 
selected sample of both single and exoplanetary microlensing events
observed towards the Galactic bulge, by taking into account the nature of the
source star: { hot, late type main sequence and cool giant stars.}
Indeed, different polarization mechanisms take place in the stellar 
atmospheres, depending on the source star type: 
photon (Thomson) scattering on free electrons, coherent (Rayleigh) scattering
off atoms and molecules, and photon scattering on dust grains, 
{ for hot, late type and cool giant stars} (with
extended atmospheres), respectively.

The analysis of the polarization curves for single-lens,
highly magnified microlensing 
events towards the Galactic bulge has shown that the polarization degree 
of the stellar light 
can reach values as high as 0.04 { percent} at the peak in the case of 
{late type source stars} 
and up to a few {percent} in the case of cool giant source stars 
(red giants) with extended envelopes. 
For these events the primary lens crosses the  source star disc 
({\it transit}
 events) and relatively large values of $P$ are thereby produced
due to large finite source effects and the large magnification
gradient throughout the source star disc.  
The time duration of the peak of 
the polarization signal may vary, from 1h to 1day, 
depending on the source star radius and the lens impact parameter.

Similar values of polarization may also be obtained
in exoplanetary events when the source star crosses the primary or the
planetary caustics. While in the former case (as for single-lens 
events) the peak of the polarization signal always occurs at symmetrical 
points with respect to
 the instant $t_0$ of maximum magnification, in the latter
case the polarization signal may occur at any (and generally unpredictable)
time during the  event.  

The natural question which arises is whether such polarization signals may be 
detectable with the present or near future technology. 
Polarimeters are nowadays available on 
large telescopes and the best possibility for measuring polarization in the
R and I bands is offered at present 
by the polarimeter in  FORS2 on ESO's VLT telescope.   
With this instrument it is possible to measure the polarization for a 12 mag
source star with a precision of 0.1 { percent} in 10 min integration time, 
and for a 14 mag star in a 1h. Indeed, for a few of the events considered
in Tables \ref{table2} and 
\ref{table4} the peak $I_0$ 
magnitude is $\simeq 12$ with the expected
maximum polarization degree lasting up to 1 day, 
thus suggesting that polarization
measurements in highly magnified microlensing events constitute 
a realistic target of opportunity for currently available instruments.

It goes without saying that polarization measurements in microlensing events
require an alert system able to predict in advance the instant
of the occurrence of the polarization peak. 
A alert system is already in operation based on OGLE
and MOA survey data \footnote{www.OGLE.astrouw.edu.pl
and www.phys.canterbury.ac.nz/MOA}, and it is particularly efficient 
in the case of highly magnified events, allowing to focus observing
resources to make intensive observations around the peak of the event.

We  emphasize that polarization measurements in
highly magnified microlensing 
events offer an unique opportunity
to probe stellar atmospheres of Galactic bulge stars.
Besides the interest related to stellar astrophysics, the 
analysis  of the polarization profile, 
which reflects that of the magnification 
light-curve, given sufficient observational precision, may in principle provide
independent constraints on the lensing parameters also for exoplanetary events.

\section*{Acknowledgments}
{ We thank the referee, R. Ignace, for useful comments and suggestions.}
We also thank H. M. Schmid and R. Gratton for useful discussions.
SCN thanks the Swiss National Science Foundation
for support during this work.


\end{document}